\documentclass[structabstract]{aa} 
\usepackage{natbib}
\bibpunct{(}{)}{;}{a}{}{,}
\usepackage{graphicx}
\usepackage{amsmath}	
\usepackage[usenames]{color}	
\usepackage{epstopdf}	
\usepackage{txfonts}
\usepackage{amssymb} 

\DeclareGraphicsExtensions{.eps,.pdf,.ps}

\newcommand{\inv} {\frac {1}}

\newcommand{\eq}[1] {Eq.\,(\ref{#1})}
\newcommand{\eqn} [1] {
\begin{equation} #1
\end{equation}}
\newcommand{\eqna} [1] {
\begin{eqnarray} #1
\end{eqnarray}}

\newcommand{\taueff}{\tau_{\rm eff}}

\newcommand{\fig}[3]{
      \begin{figure}[ht]
        \centering
         \includegraphics[width=0.98\hsize]  {#1}
        \caption{#2}
        \label{#3}
        \end{figure} }

\usepackage{txfonts}

\newcommand{\teff}{T_{\rm eff}}

\newcommand{\numax}{\nu_{\mathrm{max}}}

\newcommand{\diff}{{\mathrm{d}}}
\newcommand{\PII} {paper~II}

\usepackage[switch]{lineno} 
\modulolinenumbers[5]

\begin{document}
\title{Stellar granulation as seen  in disk-integrated intensity}
\subtitle{I. Simplified theoretical modeling}

\author{R. Samadi\inst{1}, K. Belkacem\inst{1}, H.-G. Ludwig \inst{2,3}}

\institute{LESIA, Observatoire de Paris, CNRS UMR 8109, UPMC, Universit\'e Denis Diderot, 5 Place Jules Janssen, 92195 Meudon Cedex, France
\and Zentrum f\"ur Astronomie der Universit\"at Heidelberg, Landessternwarte, Königstuhl 12, D-69117 Heidelberg, Germany
\and GEPI, Observatoire de Paris, CNRS UMR 8111, Universit\'e Denis Diderot, 5 Place Jules Janssen, 92195 Meudon Cedex, France 
}
   \offprints{R. Samadi}
   \mail{reza.samadi@obspm.fr}
   \date{\today}

  \authorrunning{Samadi et al.}
  \titlerunning{Stellar granulation as seen  in disk-integrated intensity}

   \abstract
{Solar granulation has been known for a long time to be a surface manifestation of convection. The space-borne missions CoRoT and {{\it Kepler}} enable us to observe  the signature of this phenomena  in disk-integrated intensity  on  a large number of stars.   }
{The space-based photometric measurements  show that the global brightness fluctuations and the lifetime associated with granulation obeys characteristic scaling relations. 
We thus aimed at providing  simple theoretical modeling to reproduce these scaling relations, and subsequently at inferring the physical properties of granulation  across the HR diagram.}
{We developed  a simple 1D theoretical model.  The input parameters  were extracted from 3D   hydrodynamical models of the surface layers of stars, and the free parameters involved in the model were  calibrated with solar observations. Two different prescriptions for representing the Fourier transform of the time-correlation of the eddy velocity were compared: a Lorentzian and an exponential form. Finally, we compared our theoretical prediction with 3D radiative hydrodynamical (RHD) numerical modeling of stellar granulation (hereafter  {{\it ab~initio}} approach).    }
{Provided that the free parameters are appropriately adjusted,  our theoretical model  reproduces  the observed solar granulation spectrum quite satisfactorily ; the  best agreement is obtained for an exponential form.  Furthermore, our  model results in granulation spectra that  agree  well with the {{\it ab~initio}}  approach using two 3D RHD models that are representative of the surface layers of an F-dwarf  and  a red-giant star. }
{  We have developed a  theoretical model that satisfactory reproduces   the solar granulation spectrum and gives results consistent with the {{\it ab~initio}}  approach.   The model   is used  in a companion paper as theoretical framework for  interpretating the observed scaling relations. 
}


   \maketitle
   
\section{Introduction}
\label{intro}
The solar surface  reveals irregular cellular patterns commonly called granules.  These structures  were first discovered and discussed during the 19th century \citep[see an historical review in][]{Bray67}. Their origin was first attributed by \citet{Unsold1930} to convective currents occuring beneath the visible photospheric layers, while their turbulent nature was first highlighted by \citet{Siedentopf1933}. 
Several decades later, with the advances in numerical hydrodynamical simulations, the properties of solar granulation are well explained \citep[e.g.][and references therein]{Muller99}. 

 Stars with effective temperature lower than about 7~000~K have an appreciable upper convective envelope and    are thus expected to show  -- as in the Sun ~ --  granules on their surface. 
Because  the granules  evolve with time, their evolution produces small  brightness fluctuations that 
 can now be accurately  monitored and measured  with space-based high-precision photometry measurements performed with the MOST, CoRoT and {\it Kepler} missions \citep{Matthews04,Michel08,Kallinger10b,Mathur11,Chaplin11}. 
These observations reveal that  the characteristic time $\taueff$ of the granules and the total brightness fluctuations $\sigma$ associated with them scales as a function of the frequency $\numax$ at which the solar-like oscillations peak \citep[e.g.,][]{Huber09,Kallinger10b,Chaplin11,Mathur11,Kjeldsen11}. In turn, $\numax$ is shown to scale with the cut-off frequency $\nu_c$ of the atmosphere, therefore mainly the pressure scale-height near the photosphere \citep{Brown91,Kjeldsen95,Stello09,Huber09,Mosser10,Kevin11}. 
The observed relation between the properties of the stellar granulation and $\nu_c$  were  explained on the basis of some simplified physical considerations \citep{Huber09,Kjeldsen11,Mathur11}.
A detailed  theoretical study of the observed scaling relations is, however, lacking.


A possible theoretical approach is the one proposed by \citet{Ludwig06}. This {\it ab~initio} approach consists of modeling the stellar granulation as seen in disk-integrated intensity from the intensity  emerging \emph{directly} from given 3D RHD  models.
This numerical approach was applied by several authors \citep{Svensson05,Ludwig09,Mathur11}. 
 It is very time-consuming and hence does not easily allow envisaging a large set of calculations with   different surface metal abundance, for example. Furthermore,   interpreting  the results is not trivial, and the systematic differences obtained by \citet{Mathur11}  between observed and modeled spectra of red giants are not well understood. 
 On the other hand,  a theoretical model based on a more simplified physical  approach offers the advantage of separately testing  several properties of turbulent convection. In addition, it allows one to compute  the granulation spectrum for a variety of stars on a large scale. Furthermore, it is  possible to derive   scaling relations from such a simplied theoretical model  that could provide   additional theoretical support for the observed scaling relations and extend the current theoretical scaling relations.
We here present such a simple theoretical model of the stellar granulation as seen in disk-integrated intensity.
 In the companion paper \citep[hereafter \PII]{Samadi13b}, we derive   theoretical scaling relations for $\sigma$ and $\taueff$ from this  model. 
 Comparisons with a large sample of observed stars as well as with previously published scaling relations  are reported in paper~II.

This paper I is organized as follows: 
In Sect.~2, we  outline our theoretical model for the stellar granulation spectrum  in disk-integrated intensity and  the different prescriptions adopted  for modeling the properties of turbulent granules. The free parameters introduced in the theoretical model are next calibrated in Sect.~3 such that our theoretical calculations match at best the solar observations. In Sect.~4, we compare our calculations with those obtained with  {\it ab~initio} of \citet{Ludwig06} approach using two 3D RHD models  of the surface layers of an F-dwarf star and a red giant star. Finally, Sect.~5 is devoted to the conclusion. 

\section{Modeling of the power density spectrum}

We here outline  our theoretical model for the power spectrum associated with the relative variations of the bolometric flux emerging from the star in the direction of an observer, who would measure it continuously  during a given duration (typically much longer than the timescale of the granulation). The detailed derivation of the model is presented in Appendix~\ref{thdev}, while the underlying approximations and assumptions are discussed in Appendix~\ref{approx}.
 
  
 $F(t)$ is the  bolometric flux  toward the observer at the instant $t$, and we define $\Delta F (t) =  F(t) - \left<F\right>_t$ as the instantaneous variation of the flux with respect to its time average, $\left<F\right>_t$. From now  on $\left< \right>t$ stands for a time average. The  power density spectrum (PDS) associated with the relative variation of the flux ($\Delta F (t)/\left<F\right>_t $) is hence defined by
\eqn{
{\cal F} (\nu)=  {1 \over T_0} \,  { { \left | \widehat{\Delta
        F}\right |^2  }  \over {\left<F\right>_t^2  }} \, , \label{F_nu}
}
where $T_0$ is the duration of the observation, and $\nu$ a given frequency. The operator $~\widehat{\left (~.~ \right ) }~$ is 
\eqn{
\widehat{f} (\nu) = \int_{-T_0/2}^{T_0/2} \diff t \, f (t)\, e^{i \, 2\pi
  \, \nu \, t}  \, . \label{FT}
}
We point out that with CoRoT or {\it Kepler} observations $T_0$ is typically much longer than the granule lifetime such that the  operator $~\widehat{\left (~.~ \right ) }~$ tends to the classical Fourier transform.  

We  consider  a \emph{gray atmosphere}. This is a necessary condition to obtain an analytical formulation for the granulation spectrum. 
We adopt a spherical coordinate system ($r,\theta,\phi)$ with the $z$-axis pointing toward the observer. Accordingly, the bolometric flux received from the star at the instant $t$ is  given by
\eqn{
F (t) = R_s^2 \,\int_{\phi=0}^{2 \pi} \diff \phi\, \int_{\mu=0}^1 \diff \mu \, \mu \, I (t,\tau=0, \mu,\phi) \, ,
\label{F_t}
}
where $\mu=\cos(\theta)$, $I(t,\tau,\mu,\phi)$ is the specific intensity in the direction $(\mu,\phi)$, 
 $R_s$ the radius of the star, and $\tau(r)$ the  mean optical depth 
\eqn{
\tau(r) = \int_{r^\prime=r}^{+\infty}\diff r^\prime \,  \kappa(r^\prime) \,
\overline{\rho}(r^\prime) \,  , \label{optical_depth}
}
where $\kappa$ is the   mean opacity, $\overline{\rho}$ is the mean density (i.e., averaged in time and over a sphere of radius $r$)
 and the specific intensity $I (\tau, \mu,\phi)$ is related to the source function according to \citep[e.g.,][p.~114]{Gray92}
\eqn{
I(\tau,\mu,\phi) = \int_{\tau}^{+\infty} \diff \tau^\prime \, { {e^{-\tau^\prime(r)/\mu}} \over {\mu} } \,
S(t,\tau^\prime(r),\mu,\phi) \, ,
\label{I_tau}
}
where $S$ is the source function. 


To proceed, we assume the local thermodynamic equilibrium (LTE) so that $S=B$, where $B$ is the Planck function at the instant $t$ and the position $(\tau(r),\mu,\phi)$. 
Note that LTE is fully justified because  the region where the granules are  seen most often extends to the small region around the optical depth $\tau \sim 1$ (i.e., near the photosphere).
Accordingly, and using Eqs. (\ref{F_t}) and (\ref{I_tau}), $\Delta F (t)$ is rewritten such as
\eqn{
\Delta F (t) =   R_s^2 \,\int_{\phi=0}^{2 \pi} \diff \phi\, \int_{\mu=0}^1
\diff \mu  \,  \int_0^{+\infty}  \diff \tau \,  {e^{-\tau(r)/\mu}}  
\, \Delta B \, ,\label{delta_F}
}
where we have defined $\Delta B \equiv B  - \left<B \right>_t$.

We neglect  length-scales longer than the granulation length-scales. In that case  $\Delta B$ represents the instantaneous difference  between the brightness  of the granules  situated at position $(\tau(r),\mu,\phi)$ and the brightness of the material in the steady state ($<B>_t$). 
Finally, we assume that $\kappa \rho$ varies at a length-scale longer than the characteristic size of the granule. From this set of assumptions, and after some calculations, \eq{F_nu} can be written (see details in Appendix~\ref{thdev})
\begin{align}
{\cal F} (\nu) = \int_{0}^1 \diff \mu \, \int_{0}^{+\infty} \diff \tau\, e^{-2 \tau/\mu} \, \left ( \frac{\langle {B } \rangle_t  } {F_0} \right )^2 \;  {\cal F}_\tau (\tau,\nu)
\label{F_nu_2}
\end{align}
with
\eqna{
{\cal F}_\tau (\tau,\nu) & = &  { {(2 \pi)^2 \, \kappa \rho} \over {R_s^2 \, 
    \langle {B } \rangle_t^2} }\;  \widetilde{\langle  \Delta B_1 \,  \Delta B_2  \rangle} (\nu, \vec
k=\vec 0, \tau) \; , \label{F_nu_tau}
}
 where the constant $F_0$ is given by \eq{F0}, $ \widetilde{\langle  \Delta B_1 \,  \Delta B_2  \rangle} (\nu, \vec
k, \tau)$ is the space and time Fourier transform of  $ \langle  \Delta B_1 \,  \Delta B_2  \rangle$ (see \eq{A_B}), where $k$ is a wavenumber, $\nu$ a frequency, and the subscripts   1 and 2 refer to two different spatial and  temporal positions. 


To continue we need to derive an expression for the correlation product $\langle  \Delta B_1 \,  \Delta B_2  \rangle$. To this end, we recall that $B=\sigma T^4/\pi$, where $\sigma$ is the Stefan-Boltzmann constant, and introduce $\Delta T$ as the difference between the temperature of the granule and that of the surrounding medium. We further more use the quasi-normal approximation \citep{Million41} and  assume that the scalar $\Theta \equiv \Delta T /  \langle T \rangle_t$ is isotropic and behaves as a passive scalar. $E_{\Theta}( k,\nu) $ is then introduced as its associated  spectrum \citep[Chap V-10]{Lesieur97} and  factorized into a spatial spectrum $E_\Theta(k)$ and a frequency-dependent factor $\chi_k (\nu)$ (see \eq{E_k_nu}). 
The above-mentionned set of approximations, after  tedious calculations, leads us to (see details in Appendix~\ref{thdev})
\eqn{
{\cal F}_\tau (\tau,\nu) =  \left ( { \sigma_\tau^2 \over  \nu_0   }  \right ) \, {S}_\Theta (\tau,\nu) \, , 
\label{F_nu_tau_4}
}
where 
\begin{align}
\sigma_\tau  =  \frac{12}{\sqrt{2}} \,  \sqrt { {\tau_g} \over {{\cal N}_g}  }\, \Theta_{\rm rms}^2  \, ,  \quad 
\tau_g  =  \kappa\,\rho\, \Lambda \, , \quad 
N_g   =  \frac{2 \pi R_s^2}{\Lambda^2} \, , 
\label{N_g}
\end{align}
 $\Theta_{\rm rms}$  the root-mean-square of $\Theta$ (\eq{theta_rms}), and $\Lambda$ a characteristic length (see below).
Note that in \eq{N_g}, $\tau_g$ corresponds to the  characteristic optical
thickness of the granules,  $N_g $ to the average number
of granules distributed over  half of the photosphere (i.e., at $r = R_s$), and $\sigma_\tau$ to the global brightness fluctuations
associated with the granulation spectrum  that one would see at the optical depth $\tau$. 
The RHS of \eq{F_nu_tau_4} involves the dimensionless source function ${S}_\Theta (\tau,\nu)$, whose expression is given in \eq{source}. The latter depends on $E_\Theta(k)$ and $\chi_k(\nu)$.

The adopted expression for the spatial spectrum $E_\Theta(k)$ (see \eq{eqn:E_s}) involves the length-scale $\Lambda$ as well as the  characteristic wavenumber   $k_c$, which separates the inertial-convective range from the inertial-conductive range.
Some prescriptions are required for $\Lambda$  and $k_c$, however. We hence assume   that $\Lambda  =  \beta \, H_p$, where $H_p$ is  the pressure scale height and $\beta$  a free parameter.
For  $k_c$, we adopt the prescription $ k_c =  \zeta  \, \left ( { {\epsilon} \over  {\chi_{\rm rad}^3} }  \right )^{1/4}$, where $\zeta$ is a free parameter introduced  to exert some control on this prescription, $\chi_{\rm rad}$ is the radiative diffusivity coefficient, and $\epsilon$ is the rate of injection of kinetic energy into the turbulent cascade.

We turn now to the frequency component $\chi_k(\nu)$. In a strongly turbulent medium,  $\chi_k$ is  well described by a Lorentzian function
\begin{equation} 
\chi_k(\nu) = {1 \over {\pi \nu_k}} \, \frac{\displaystyle{1}}{\displaystyle{1+\left( \nu/\nu_k \right)^2}} \; ,
\label{LF}
  \end{equation} 
where $\nu_k$ is by definition the half-width at half-maximum of $\chi_k(\nu)$. The expression for $\nu_k$ is given in \eq{nu_k}, which involves the free parameter  $\lambda$. This was introduced  to have some control on this definition. As an alternative for a Lorenztian function (see Appendix~\ref{Frequency spectrum}), we also consider   an exponential form 
\begin{equation} 
\chi_k(\nu) = { {\ln 2 } \over {2 \, {\nu_k}}} \, \exp \left [ - \left | \ln 2 \, { \nu \over \nu_k } \right |  \right  ]  \, .
\label{EF}
  \end{equation}

\section{Calibration using solar observations}
\label{sun}

The theoretical granulation representative of the Sun were computed on the basis of the present model using inputs extracted from a 3D hydrodynamical model 
 of the solar surface layers.  To reproduce the solar observations, we calibrated the free parameters involved in the theory. 

\subsection{Computation details}
\label{method}

 We computed the theoretical PDS of the granulation ($\cal F$) according to Esq.~(\ref{F_nu_2}), (\ref{F_nu_tau_4}), and (\ref{N_g}) together with \eq{F0} and the set of Eqs.~(\ref{source})--(\ref{K_xi}). The different quantities involved in the theoretical model (stratification, convective velocity, etc)  were obtained from a 3D model
 in a similar way as in \citet{Samadi02I}. However, while \citet{Samadi02I} extracted these quantities from horizontal averages at constant geometrical depth, we here performed  the averages at a constant optical depth $\tau$. This is justified because   the RHS of \eq{F_nu_2} is integrated over the optical depth. 

For later use, we define $\sigma$ as the root-mean-square (rms) brightness fluctuations associated with a given PDS. The latter satisfies the following relation
\eqn{
\sigma ^2 =  \int_{-\infty}^{+\infty} \diff \nu \; {\cal F} (\nu) \, ,
}
where ${\cal F} (\nu)$  refers to a given PDS.
Another important characteristic of the granulation spectrum is its associated timescale $\taueff$, which is here defined, following  \citet{Mathur11}, as the e-folding time  associated with the auto-correlation function (ACF) of the relative flux variations caused by the granulation. 
Different PDS are next compared, in terms of their shape (i.e., $\nu$ variation),   $\sigma$ and   $\taueff$. 

We considered a 3D model representative of the surface layers of the Sun \citep[see details in][]{Ludwig09} and computed the associated theoretical  PDS. This was then compared with the PDS  obtained from the green channel of the SOHO/VIRGO three-channel sun-photometer \citep[SPM, see][]{Frohlich97}. We  multiplied  the original data by the instrumental function response function as derived by \citet{Michel09}  to convert the observed photometric fluctuations  in terms of bolometric ones. 
To compare theoretical PDS with the observations, we systematically added the instrumental whited noise component, which we measured at high frequency on the observed solar spectrum.

\subsection{Lorentzian   versus exponential $\chi_k$}

\label{LF_vs_EF}

We first assumed by default $\lambda=1$, $\beta=1$ and $\zeta=1$ and adopted a Lorentzian shape for $\chi_k$ (\eq{LF}). The  theoretical PDS underestimates $\sigma$  by a factor of about 30  and  the width of the solar granulation by a factor about five (not shown). 
Part of this significant discrepancy is  due to our prescription for $\Lambda$ (\eq{lambda_v}), $k_c$ (\eq{k_c}), and $\nu_k$ (\eq{nu_k}). The estimation of these quantities can been controlled by the  parameters $\beta$ and $\zeta$, and  the product $\beta \, \lambda$, respectively.
 However, there is some degeneracy between $\zeta$ and $\beta$ (see below). At fixed value of $\zeta$, we then  simultaneously adjusted  $\beta$ and $\beta \, \lambda$ such that  $\sigma$ and the width of the spectrum best matches  the observations. 
 This led to  $\beta=14.8$ and $\lambda \beta=3.7$.
The resulting  theoretical PDS  is shown in Fig.~\ref{pds_sun3d}. We obtained an overall satisfactory agreement between the theoretical PDS and the observations, except for the frequency variations  at high frequencies.

\fig{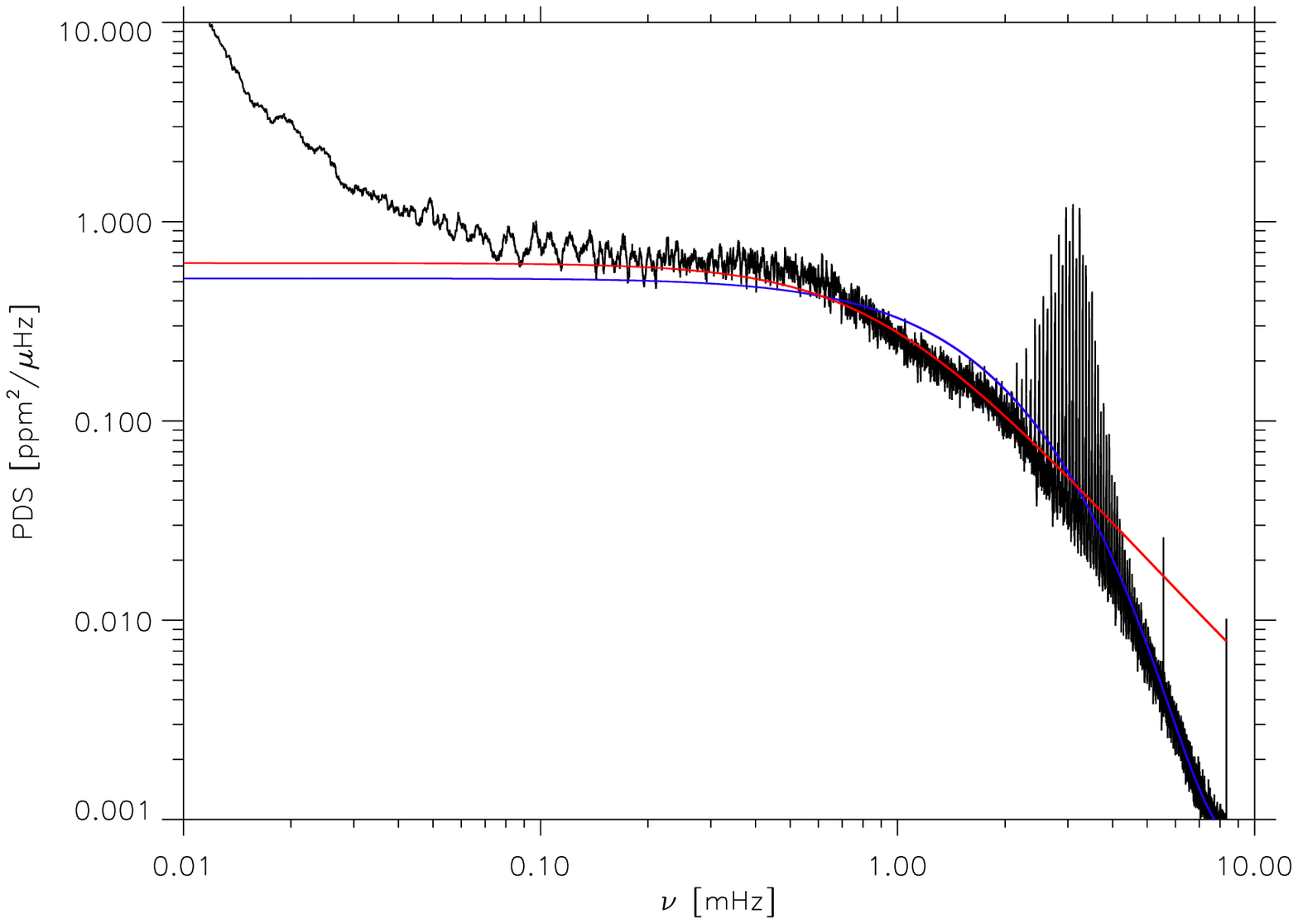}{Power density spectrum in ppm$^2/\mu$Hz as a function of frequency $\nu$. In black: PDS  obtained from the green channel of  the SOHO/VIRGO three-channel sun-photometer. 
In red and blue: theoretical PDS of the granulation background 
 ($\cal F$) computed using the quantities obtained from the solar 3D model. The red curve assumes a Lorentzian function for $\chi_k$, the blue curve an exponential functional. 
 }{pds_sun3d}

There are several pieces of observational evidence that the granulation spectrum significantly departs at high frequencies from  a Lorentzian function \citep[e.g.,][and references therein]{Mathur11}. Furthermore, calculations performed with  the \citet{Ludwig06} method  also confirm this general trend \citep{Ludwig09,Ludwig11}.
Accordingly, we now considered the exponential function given by \eq{EF}. The free parameter $\beta$ and the product $\beta \, \lambda$ were adjusted such that the theoretical PDS reproduces  the observations best. For this we fitted the observed solar spectrum by means of the maximum-likelihood estimator \citep[see e.g.][]{Toutain94,Appourchaux98}.
 For $\zeta=1$, the fit leads to  $\beta = 7.8$ and $\beta \,\lambda  = 3.0$. 
As seen in Fig.~\ref{pds_sun3d}, we obtained an overall satisfactory agreement with the observations. In particular,  the observations are much better fitted at high frequencies than when a  Lorentzian function is adopted. 
Consequently, unless mentioned otherwise, we adopted the  exponential function from now on.

Other values of the parameter $\zeta$ were also tested. 
 For each adopted value of $\zeta$, we have adjusted  the parameter $\beta$ and the product $\beta \, \lambda  $ such as to reproduce the  solar data best. They all result in almost the same agreement with the solar observations. As for the Lorentzian $\chi_k$ (see Sect.~\ref{LF_vs_EF}), there is thus a degeneracy between the parameters $\beta$ and $\zeta$. Therefore, the observed solar granulation background does not permit one to constrain these parameters independently. However, the granules observed on the solar surface have a typical size of about 2~Mm \citep[see e.g.][]{Muller89,Roudier91}. Furthermore, the observations reported by \citet{Espagnet93} and \citet{Hirzberger97}  suggest that $k_c/k_0 \approx 2$. These observations accordingly favor a value of $\zeta$ of about five  since this value results in $k_c/k_0 = 1.9$ and $\Lambda=1.7$~Mm. In the following we  therefore assumed $\zeta=5$.  The corresponding value of $\beta$ is 12.3, which value  is within the range found by \citet{Trampedach13}, that is, 9-13.

\section{Comparison with  the \citet{Ludwig06}  {\it ab~initio} modeling}
\label{comp_direct}

We compare here theoretical PDS computed with calculations  performed on the basis of the \citet{Ludwig06} {\it ab~initio} modeling assuming   the same 3D hydrodynamical models in both cases.
 Two different 3D models were considered: a 3D model representative of the surface layers of an F-type main-sequence star (Sect.~\ref{F3Dmodel}) and a second representative of the surface layers of a red giant star (Sect.~\ref{RG3Dmodel}). These 3D models constitute  two extreme cases in the H-R diagram. 

Because calculation of the PDS requires  knowing  the stellar radius $R_s$, this was obtained by matching a complete 1D standard model to the stratification of the 3D model \citep[see details in][]{Trampedach97,Samadi08}.  The standard  1D model was   computed using the CESAM2K code \citep{Morel08}.

\subsection{F-type main-sequence stellar model}

\label{F3Dmodel}

\begin{figure}[ht]
  \centering
        \includegraphics[width=0.8\hsize]  {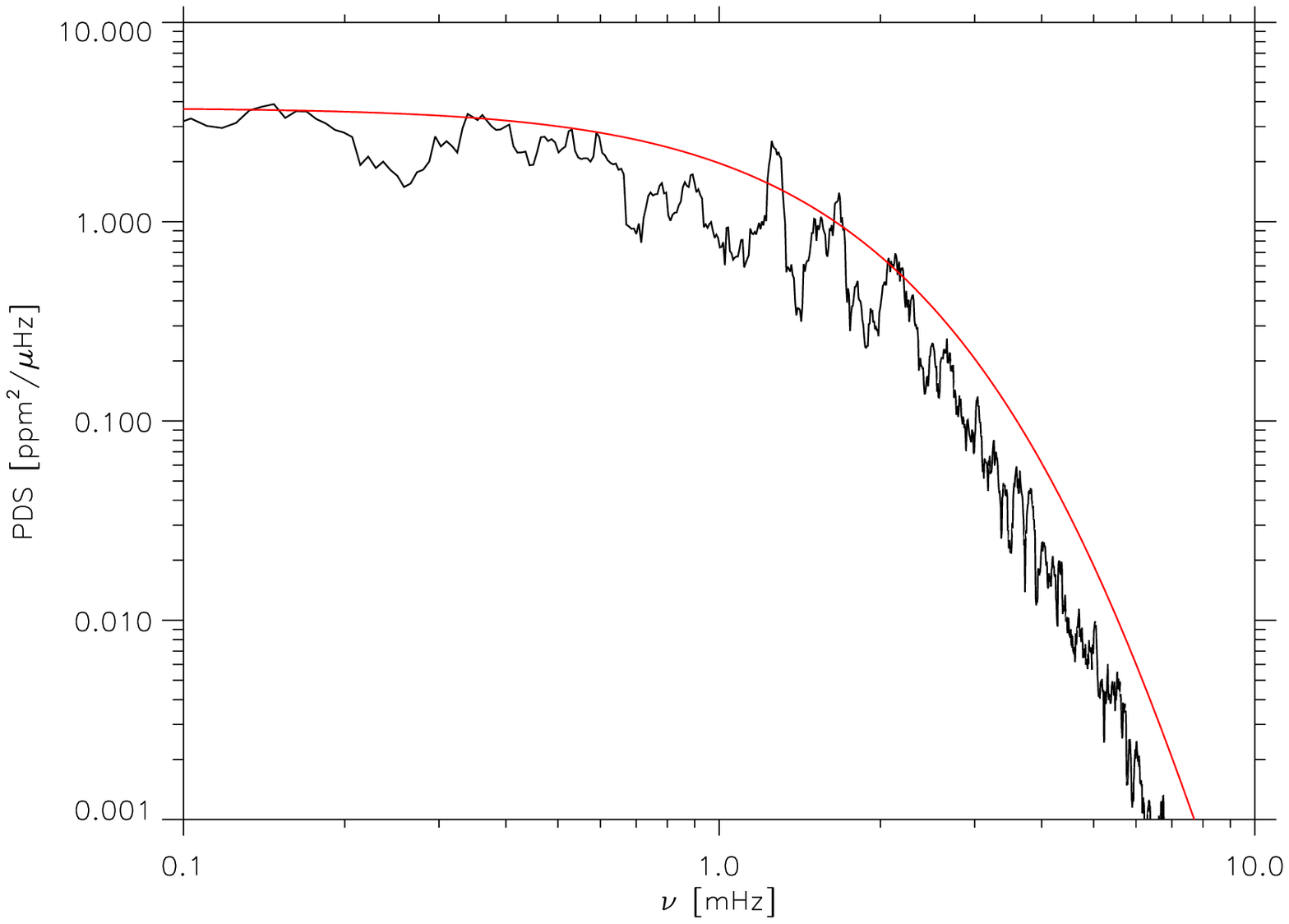} 
\includegraphics[width=0.8\hsize]  {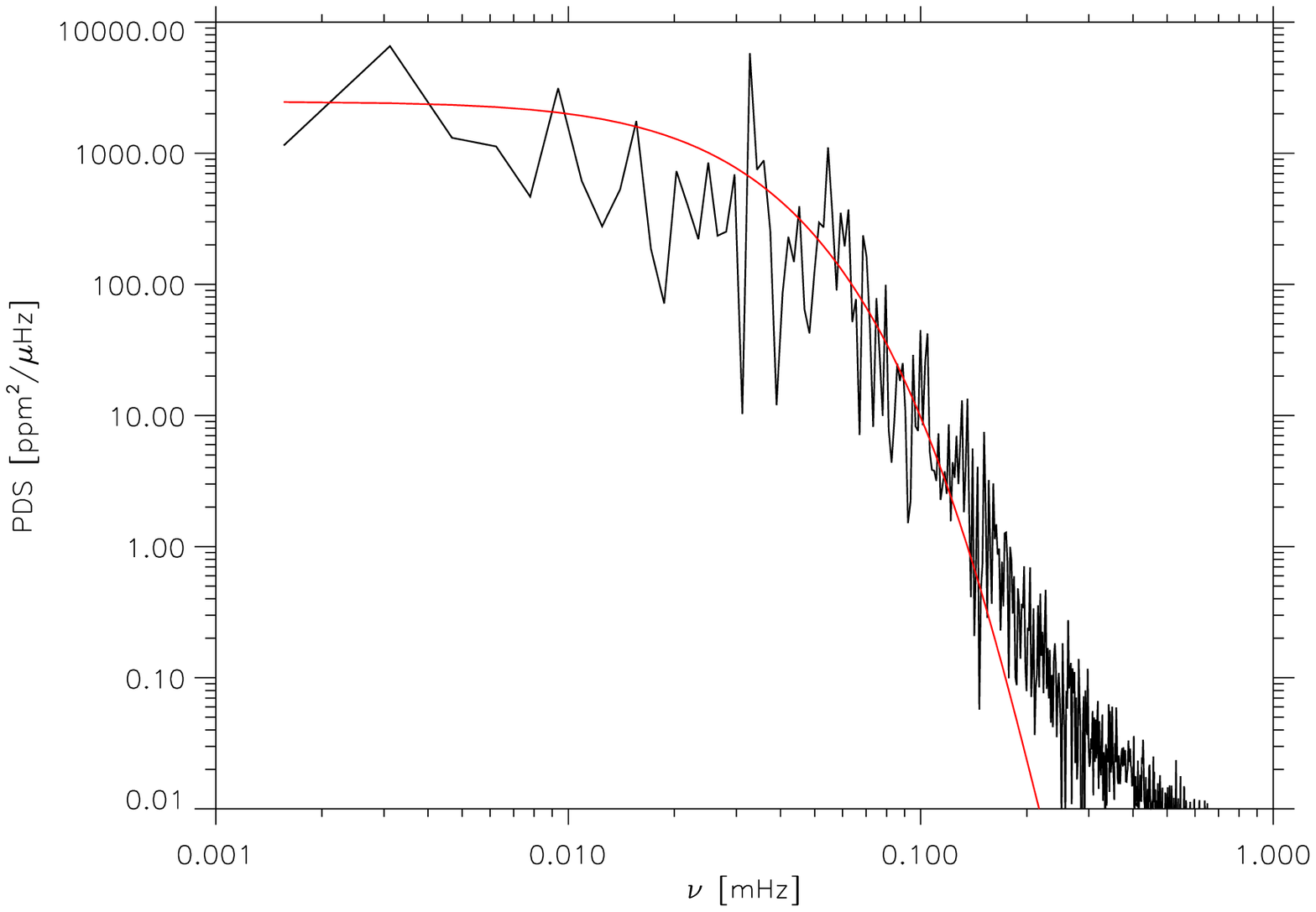}
       \caption{  {\bf Top:} Power density spectrum in ppm$^2/\mu$Hz as a function of frequency $\nu$ obtained for a 3D RHD  model representative of an F-dwarf star (see Sect.~\ref{F3Dmodel}). The black curve corresponds to the theoretical PDS obtained by  \citet{Ludwig09} on the basis of the {\it ab~initio} approach, the red curve to the PDS obtained with our theoretical model  using inputs extracted from the same  3D  model (see Sect.~\ref{method}). {\bf Bottom:} Same as the top panel for the 3D model representative of a red-giant star (see Sect.~\ref{RG3Dmodel}).   }
       \label{pds_3D}
       \end{figure}

\citet{Ludwig09} have computed two F-dwarf 3D models that included the CoRoT target HD~49933 in terms of surface metal abundance.  Of these two 3D models,  we considered here the one with the solar surface abundance. 
This model has  $\teff=$ 6725~K  and $\log g=4.25$. 
We computed the theoretical PDS as detailed in Sect.~\ref{method}.
 We adopted the values of the free parameters that give  the best agreement between the model and the observations for the Sun (see Sect.~\ref{sun}).  The result of the calculation is presented in  Fig.~\ref{pds_3D} (top panel).

We  compared the theoretical PDS with the spectrum obtained by  \citet{Ludwig09} on the basis of the {\it ab~initio} method.  Their calculation was based on a radius of $R_{S,Ludwig}= 1.35\,R_\odot$, while our associated global 1D  model has a radius of $R_S=1.473\,R_\odot$. To compare ours with  their results, we multiplied their spectrum by $(R_{S,Ludwig}/R_S)^2$ since the PDS is inversely proportional to the square of the radius \citep{Ludwig06}.
The resulting  PDS is shown in Fig.~\ref{pds_3D} (top panel).

The $\nu$-variation of the PDS obtained with the {\it ab~initio} approach is  well reproduced. 
For the characteristic time  $\taueff$, our theoretical calculations result in a value of $\taueff$ 5\,\% lower than for \citet{Ludwig09}. 
The rms of the global brightness fluctuations associated with their spectrum is  $\sigma= 84 \pm 17$~ppm. Our theoretical model results in an rms brightness fluctuation of $\sigma= $97.4~ppm. This is about $16$\,\% higher than  the result of \citet{Ludwig09}.
 It is  worth mentioning that,  consistently with the \citet{Ludwig09} results, our model also results  in a significant overestimation of the measured $\sigma$ for this F-dwarf star.  Furthermore, we  show in \PII\ that our theoretical model results for  F-dwarf stars in a systematic overestimation of $\sigma$. As discussed in \PII, this trend  is very likely  linked to the impact of the high level of magneticactivities on the surface convection \citep[see also][]{Chaplin11b,Chaplin11}. We therefore emphasize that all our calculations must be rigorously considered to be   valid only for stars with a low activity level.

\subsection{Red giant model}
\label{RG3Dmodel}

We  considered  a 3D model of the surface layers of a red-giant  star characterised by  $\log g=2.5$ and  $\teff = 4964~\pm~22$~K \citep[see details in][]{Ludwig11}. The associated theoretical PDS is compared in Fig.~\ref{pds_3D} (bottom panel) with the PDS computed by \citet{Ludwig11}  on the basis on the  \citet{Ludwig06} {\it ab~initio} approach. Their calculation assumed a radius $R_{S,Ludwig}= 10\,R_\odot$, while our  associated global 1D  model has a radius of $R_s=17.8\,R_\odot$. Accordingly, we  multiplied their PDS by $(R_{S,Ludwig}/R_s)^2$. 

Except at very high frequency ($\nu \gtrsim 150~\mu$Hz), we obtain a good match between the two calculations, both in terms of amplitude and $\nu$-variation.
Our theoretical spectrum results in $\sigma = 352~$ppm, which  is only 9\,\% higher than the one obtained by \citet{Ludwig11}. 
For $\taueff$, we obtain $\taueff= 1.01 \times 10^{4}$~s. This is 29\,\% higher  than the value derived from the \citet{Ludwig11} PDS. 

We note that  assuming  different sets of the calibrated parameters  (see Sect.~\ref{LF_vs_EF}) has a negligible impact on the $\nu$-variation of the PDS and a moderate effect on $\sigma$, for both the red giant and F-dwarf models.  
Furthermore, for both 3D models, assuming a Lorenztian function for $\chi_k(\nu)$ instead of an exponential function, results in a considerable discrepancy at high frequency, as in the case of the Sun (not shown).

\section{Conclusion}
\label{conclusion}

We have developed a simple 1D theoretical model for the granulation spectrum. One advantage of this model is that it can be  used to compute theoretical  granulation spectra on a large scale.  Another benefit is that  any prescription for the turbulent spectrum can been considered, in particular, different prescriptions can be tested for $\chi_k$, the Fourier transform  of the time correlation function  of the eddy velocity  at a fixed  $k$ wave-number.  In this way, we have established a link between the $\nu$-variation of the granulation spectrum in intensity with the  frequency component $\chi_k$. 

The theoretical model  was first  applied to the solar case. 
The stratification of the surface layers of the Sun and  the properties of its surface convection were obtained from a 3D hydrodynamical model of the surface layers of the Sun. 
Assuming default values for the free parameters involved in the theoretical model, our theoretical model strongly underestimates the quantities $\sigma$ and $\taueff$ derived from the observed solar granulation spectrum. 
Nevertheless, the solar granulation spectrum can be well reproduced  by adjusting two of the three free parameters involved in the theoretical model, while the degeneracy between the free parameters was removed by using constraints from solar images.

Two different functions for $\chi_k(\nu)$ were tested: a Lorentzian function and an exponential one.
 Whatever the choice of the free parameters, the $\nu$-variation of the solar spectrum is not reproduced at high frequencies with the Lorentzian function.  Indeed, at high frequencies the theoretical spectrum decreases as $\nu^{-2}$, while the observations decrease much more rapidly with $\nu$.
 On the other hand, adopting an exponential  $\chi_k$ results in a much better agreement  at high frequency.
As discussed in Appendix~\ref{choice for chi}, this result confirms the previous claim by \citet{Nordlund97} that the granules seen in the visible part of the atmosphere have a low level of turbulence.

We then compared theoretical PDS with PDS calculated on the basis of the \citet{Ludwig06} {\it ab~inito} approach, that is from the timeseries of the intensity emerging directly from 3D hydrodynamical models of the surface layers of stars.
 The time-averaged properties of the surface layers were obtained from two 3D hdyrodynamical models, one representative of an F-type star and the second one for a red giant star. 
The theoretical PDS reproduces the main characteristics of the PDS obtained with the {\it ab~initio} modeling. As in the solar case, the best agreement was obtained with an exponential $\chi_k(\nu)$.

 However, some residual differences remain between our theoretical calculations  and the solar granulation spectrum  as well as between our calculations and those obtained with the {\it ab~initio}  approach.  
As discussed in Appendix~\ref{residual}, all these differences  do not exhibit a severe defect of our theoretical model, however, and we conclude   that  the main physical assumptions of the theoretical model are validated by our different comparisons. 
Further theoretical developments are required to improve the  modeling of the granulation background, however.

We  compute in the companion paper on the basis of this theoretical  model the global brightness fluctuations ($\sigma$) and the lifetime ($\taueff$) associated with the stellar granulation for a set of stars with different spectral types and luminosity classes. 
This will then allow us to derive theoretical scaling laws for  $\sigma$ and $\taueff$, which we then compare with {\it Kepler} observations.

\begin{acknowledgements}
SOHO is a mission of international collaboration between ESA and NASA. RS and KB acknowledge financial support from the Programme National de Physique Stellaire (PNPS) of CNRS/INSU and from Agence Nationale de la Recherche (ANR, France) program  ``Interaction Des \'Etoiles et des Exoplan\`etes'' (IDEE, ANR-12-BS05-0008).  HGL acknowledges financial support by the Sonderforschungsbereich SFB\,881
``The Milky Way System'' (subproject A4) of the German Research Foundation (DFG). RS thanks Fr\'ed\'eric Baudin for useful discussions about statistical issues.
\end{acknowledgements}

\bibliographystyle{aa}


%
%
\Online
\appendix
\section{Theoretical developments}
\label{thdev}

\subsection{General expression of the PDS}
\label{appendix:general_expression}
We establish here our general expression of the PDS, that it is  Eqs. (\ref{F_nu_2}) and (\ref{F_nu_tau}). We start from the term of \eq{delta_F}, which enters in the RHS of \eq{F_nu}. 
We  apply the operator given by \eq{FT} on \eq{delta_F}, it gives
\eqna{
\widehat{\Delta F} (\nu) & = & R_s^2 \, \int \frac{\diff^3{\vec r}}{r^2} \, \int_{-T_0/2}^{T_0/2} \diff t \; \rho  \kappa \,  {e^{-\tau(r)/\mu}}  \,  e^{i \, 2\pi \, \nu \, t}  \, \Delta B(t,\vec r)  \, ,
\label{delta_F_3}
}
where the space integration is performed over half of the (stellar) sphere,  $\diff^3r = - \diff \mu \, r^2 \, \diff r \, \diff \phi $,  and $\diff \tau = \kappa \, \rho$.

 The extent of the region where the stellar granulation is seen is very small compared to the stellar radius $R_s$, such that $r$ is almost constant over the optical depth range relevant for calculating the flux. Therefore, in very good approximation we have  $\left ( R_s / r \right) ^2 \simeq 1 $ in the integrand of \eq{delta_F_3}.  Accordingly, \eq{delta_F_3} can be simplified as
\eqna{
\widehat{\Delta F} (\nu) & = & \int \diff^3{\vec r} \, \int_{-T_0/2}^{T_0/2} \diff t \; \rho  \kappa \,  {e^{-\tau(r)/\mu}}  \,  e^{i \, 2\pi \, \nu \, t}  \, \Delta B(t,\vec r)  \, .
\label{delta_F_4}
}
Since $\widehat{\Delta F} (\nu)$ cannot be evaluated in a deterministic way but  a statistical one only, we  consider the square of \eq{delta_F_4} and average it over a large number of independent realizations. This gives
\begin{eqnarray}
 \langle  |  \widehat{\Delta F}  |^2  \rangle (\nu) &  =  & \int \diff^3{\vec r_1} \,  { \kappa(r_1) \rho(r_1)} \,  {e^{-\tau(r_1)/\mu (\vec r_1) }}  \,
\int_{t_1=-T_0/2}^{T_0/2} \diff t_1 \,  e^{i \, 2\pi \, \nu \,
  t_1}   \label{delta_F_5}\nonumber  \\ 
&& \times \,  \int \diff^3{\vec r_2} \, { \kappa(r_2) \rho(r_2)}  \,  {e^{-\tau(r_2)/\mu (\vec r_2) }} \, \int_{t_2=-T_0/2}^{T_0/2} \diff t_2 \,  e^{-i \, 2\pi \, \nu \, t_2}\nonumber   \\ 
&& \times  \, \langle  \Delta B_1 \,  \Delta B_2  \rangle \, ,
\end{eqnarray}
where $\Delta B_1$ (resp. $\Delta B_2$) corresponds to the quantity $\Delta B$ evaluated at the time-space position $(t_1,\vec r_1)$ (resp.  $(t_2,\vec r_2)$).

We then define the following new coordinates
\begin{align}
t_0 = \left ( t_1 + t_2 \right )/2 &\quad t^\prime=t_2-t_1 \, , \label{t_0}\\
\vec r_0 = \left ( \vec r_1 + \vec r_2 \right )/2 &\quad \vec r=\vec
r_2-\vec r_1 \, . 
\label{r_0}
\end{align}
 In these coordinates, $\vec r $ and $t^\prime $ are the spatial correlation and temporal correlation lengths  associated with the local properties of the turbulence, while $\vec r_0$  and $t_0$ are mean  space and time positions.   
Using the new coordinates given by Eqs.(\ref{t_0}) and (\ref{r_0}) in \eq{delta_F_5} leads to 
\begin{align}
\label{delta_F_5}
\langle  |  \widehat{\Delta F}  |^2  \rangle (\nu)
& =  2 \pi \int  \int_{-T_0/2}^{T_0/2}   \diff t_0 \, 
  \diff^3{\vec r_0} \\ \nonumber 
&\times    { \kappa(r_0) \, \rho(r_0)\,  e^{-2 \tau_0/\mu_0}}\,   \Gamma (t_0,\vec r_0, \nu)  \, , 
\end{align}
where 
\begin{align}
\Gamma(t_0,\vec r_0,\nu ) &=    \int \diff^3{\vec r} \;
\frac{  \gamma(\vec r_0, \vec r)}{2\pi} \,  \int_{T_0/2}^{T_0/2} \diff t^\prime \, 
  e^{- i \, 2\pi \, \nu \, t^\prime} \,  \langle  \Delta
B_1 \,  \Delta B_2  \rangle  \\ 
\gamma( \vec r_0, \vec r) &= 
\frac{ \kappa(r_1) \rho(r_1) \,  \kappa(r_2) \, 
  \rho(r_2)}{ \kappa(r_0) \rho(r_0)} \,  e^{ 2\tau_0/\mu_0
  - \left ( \tau_1/\mu_1 + \tau_2 /\mu_2 \right ) } 
  \label{gamma}
\end{align}
and $\tau_i = \tau(r_i), \;  \mu_i = \mu(\vec r_i)$ with $i=\{0,1,2\}$. 

At this stage, a tractable expression for \eq{delta_F_5} calls for more assumptions. We assume that $T_0$ is much longer than the granule lifetime. 
 We neglect  length-scales longer than the granulation length-scales. In that case  $\Delta B$  represents the instantaneous difference  between the brightness  of the granules  situated at the position $(\tau(r),\mu,\phi)$ and the brightness of the material in the steady state ($<B>_t$).
Accordingly,  \eq{delta_F_5} reduces to 
\begin{align}
\langle  |  \widehat{\Delta F}  |^2  \rangle & = 
\left ( 2 \pi \right )^2 \,  R_s^2 \, T_0 \, \int_{0}^1 \diff \mu_0
\,  \int_{0}^{+\infty} \diff \tau_0  \, {e^{-2 \tau_0/\mu_0}}  \,
\Gamma (\tau_0,\nu) \label{delta_F_6} \\
\Gamma(\tau_0,\nu)  & =   \int \diff ^3 \vec r  \, \frac{\gamma(r_0,r)}{2 \pi }  \int_{-\infty}^{+\infty} 
e^{- i \, 2\pi \, \nu \, t^\prime} \,   \langle  \Delta B_1 \,  \Delta B_2  \rangle \, \diff t^\prime 
 \label{Gamma_2} \\
\gamma(r_0,r) & =  \frac{ \kappa(r_1) \rho(r_1) \,  \kappa(r_2)
  \rho(r_2)}{ \kappa(r_0) \rho(r_0)} \label{gamma_2} \, .
\end{align}

We now assume that $\kappa \rho$ varies at a length-scale longer than the characteristic size of the granules. This assumption is discussed in Appendix~\ref{approx}. Accordingly, $\gamma \simeq \kappa(r_0) \rho(r_0)  $ and $\Gamma$ reduces to
\begin{align}
 \Gamma(\tau_0,\nu)   & =   (2 \pi)^2 \, { {\kappa (\tau_0) \rho (\tau_0)} }\, \widetilde{\langle  \Delta B_1 \,  \Delta B_2  \rangle} (\nu,\vec k = \vec 0) \, , 
\label{Gamma_3}
\end{align}
where $\widetilde{\langle  \Delta B_1 \,  \Delta B_2  \rangle}$ is the space and time Fourier transform of $ \langle  \Delta B_1 \,  \Delta B_2  \rangle$, defined as 
\begin{align}
\widetilde{\langle  \Delta B_1 \,  \Delta B_2  \rangle} (\nu,\vec k) \equiv \inv{ (2\pi)^3} \int
\diff t^\prime \int \diff^3 r \, e^{- 2 i  \pi \nu t^\prime - i \vec k .\vec r}
\,  \langle  \Delta B_1 \,  \Delta B_2  \rangle  \, .
\label{A_B}
\end{align} 
Note that, from now on, we substitute the notation $\tau_0$ by $\tau$. 

 We now turn to the time-averaged flux $\langle F \rangle_t$. From Eqs. (\ref{F_t}), (\ref{optical_depth}), and (\ref{I_tau}) one derives
\eqn{
\langle F \rangle_t = 2 \pi \, R_s^2 \, F_0 \; ,
\label{F_mean}
}
where we have defined
\eqn{
F_0  \equiv   \int_{0}^1 \diff \mu \, \int_{0}^{+\infty} \diff \tau\, e^{- \tau/\mu} \,   \langle {B } \rangle_t (\tau)   \, .
 \label{F0}
}
Finally,  using \eq{delta_F_6},  \eq{Gamma_3} and \eq{F_mean} we  derive the general expressions of Eqs.~(\ref{F_nu_2}) and (\ref{F_nu_tau}). 
 The term  ${\cal F}_\tau$ (\eq{F_nu_tau})  stands for the PDS of the granulation as it would be seen at the optical depth $\tau$.  However, the \emph{observed} PDS associated with the granulation background is given by \eq{F_nu_2} and corresponds to the sum of the spectra seen at different layers in the atmosphere, but weighted by the term $e^{-2 \tau/\mu}\, \left ( {\langle {B } \rangle_t   } / {  F_0 } \right ) ^2 $. 

\subsection{Source function ($\langle  \Delta B_1 \,  \Delta B_2  \rangle$)}

\label{appendix:source_function}

To proceed we need to derive an expression for the correlation product $\langle  \Delta B_1 \,  \Delta B_2  \rangle$. To this end, we recall that $B=\sigma T^4/\pi$, where $\sigma$ is the Stefan-Boltzmann constant, and introduce $\Delta T$ as the difference between the temperature of the granule and that of the surrounding medium. We thus have 
\eqn{
\Delta B  = \left  (  \left (1 + \Theta  \right ) ^4 -1  \right ) \,  \langle B \rangle_t \; , 
\label{delta_b}
}
where $\Theta \equiv \Delta T /  \langle T \rangle_t $. 
The second-order  Taylor expansion of the RHS of \eq{delta_b} gives
\eqn{
\Delta B    =  \left ( 4 \, \Theta  +  6  \, \Theta ^2   \right ) \, \langle B \rangle_t  \, . 
\label{delta_b_2}
}
 We have neglected the third and fourth order terms in $\Theta$. Indeed, a 3D hydrodynamical simulation of the solar surface shows that terms higher than second order contribute  less than about 15\,\%  of \eq{delta_b}. 
Accordingly,
\begin{align}
\label{A_B_2} 
\langle & \Delta B_1 \,  \Delta B_2  \rangle 
= \\ \nonumber
&4 \left ( 4 \, \langle \Theta_1 \,
  \Theta_2  \rangle + 9  \langle \Theta_1^2  \, \Theta_2^2 \rangle +
  6  \langle \Theta_1 \,  \Theta_2^2 \rangle  + 6  \langle \Theta_2
  \,  \Theta_1^2 \rangle   \right ) \, \langle B \rangle_t^2   \, ,
\end{align}
where $\Theta_1 \equiv \Theta (\vec r_1,t_1))$ and  $\Theta_2 \equiv \Theta (\vec r_2,t_2)$.

We  now adopt the quasi-normal approximation (QNA). This approximation is rigorously valid for normally distributed quantities.
Departure from this approximation is discussed in Appendix~\ref{approx}.  Normally distributed  quantities are necessarily symmetric, such  that  $\langle \Theta_1 \, \Theta_2 ^2 \rangle =0$ and  $\langle \Theta_2 \, \Theta_1 ^2 \rangle =0$. This approximation also implies \citep[e.g.,][Chap. VII-2]{Lesieur97}
\eqn{
\langle \Theta_1^2  \, \Theta_2^2 \rangle = \langle \Theta_1^2 \rangle  \, \langle \Theta_2^2 \rangle + 2 \langle \Theta_1 \Theta_2 \rangle ^2 \; . 
\label{qna}
}
The first term in the RHS of \eq{qna} does not contribute in the time Fourier domain, except at the null frequency. Accordingly,
\begin{align}
\langle & \Delta B_1 \,  \Delta B_2  \rangle =  \left[ 16 \, \langle \Theta_1 \Theta_2 \rangle +  72  \langle \Theta_1 \Theta_2 \rangle^2   \right] \,  \langle B \rangle_t^2 \; .
\label{A_B_3}
\end{align}
Now using the Parseval-Plancherel relation, \eq{A_B_3} becomes
\begin{align}
\label{A_B_4}
\widetilde{\langle  \Delta B_1 \,  \Delta B_2  \rangle} (\nu,\vec k) =    \langle B \rangle_t^2  \left[ 16  
\widetilde{ \langle \Theta_1 \Theta_2 \rangle} (\nu,\vec k)  + 72 \,  \widetilde{\cal B}_\Theta (\nu,\vec k) \right] \; , 
\end{align}
with
\eqna{
\tilde {\cal B}_\Theta (\nu,\vec k)  \equiv  &  \int \diff \nu^\prime  \int  \diff^3k^\prime 
   \widetilde{ \langle \Theta_1 \Theta_2 \rangle} (\nu^\prime,\vec k^\prime)  \widetilde{ \langle \Theta_1 \Theta_2 \rangle} (\nu^{\prime\prime}, \vec k^{\prime\prime}) \label{B_theta} \; ,
}
where $\nu^{\prime\prime}=\nu+\nu^\prime$ and $\vec k^{\prime\prime}=\vec k+\vec k^\prime$.

We assume that the scalar $\Theta$ is isotropic. Accordingly,  
 the spatial Fourier transform of $\langle \Theta_1 \Theta_2 \rangle (r,\nu)$ is given by \citep[see e.g.][Chap V-10]{Lesieur97}
\begin{align}
\widetilde{ \langle \Theta_1 \Theta_2 \rangle}(\vec k,\nu) = \frac {E_\Theta (k,\nu) } {2 \pi k ^ 2}  \; . 
\label{A_theta}
\end{align}
The scalar spectrum $E_{\Theta}( k,\omega) $ is  related 
to the scalar variance as  \cite[Chap V-10]{Lesieur97}
\begin{align}
\frac{1}{2}  \langle \Theta^2 \rangle (\tau) = {1 \over 2} \Theta_{\rm rms}^2 
=  \int_{-\infty}^{\infty} \diff \nu  \int_0^{\infty} \diff k \;  E_{\Theta}(k,\nu)  \; , 
\end{align}
where   $\Theta_{\rm rms}$ is by definition the rms of $\Theta$, which is related to $\Delta T_{\rm rms} $  (the rms of $\Delta T$) according to
\eqn{
 \Theta_{\rm rms }^2 = {\Delta T_{\rm rms}^2 \over  \langle T \rangle_t^2 } \label{theta_rms} \, .
} 
Following \citet{Stein67}, the scalar energy spectrum $ E_\Theta(k,\nu) $ can be factorized into a 
spatial spectrum $E_\Theta(k)$ and a frequency-dependent factor $\chi_k (\nu)$  according to
\eqn{
E_\Theta( k,\nu) =E_\Theta(  k) \, \chi_k(\nu) \; ,
\label{E_k_nu}
} 
where the  frequency-dependent factor $\chi_k (\nu)$ is normalized such that
\eqn{
\int_{-\infty}^{+\infty} d\nu\, \chi_k (\nu) = 1 \;. 
}

\subsection{Final expression of relative flux variations}

 The final expression for the term ${\cal F}_\tau $ (\eq{F_nu_tau}) that appears in the RHS of  \eq{F_nu_2} is  obtained by 
inserting  \eq{A_B_4} to (\ref{A_theta}) into \eq{F_nu_tau}, so that
\eqn{
{\cal F}_\tau (\tau,\nu) =   {{(2\pi)^2 \, \kappa \rho } \over R_s^2}   \;  \left[   16  \, \widetilde{ \langle \Theta_1 \Theta_2 \rangle} 
  + 72 \,  \tilde{\cal B}_\Theta 
\right] \, , 
\label{F_nu_tau_2}
} 
where the two terms in the RHS of \eq{F_nu_tau_2} are considered for $ (\tau,\nu,k=0)$. 

Equation (\ref{F_nu_tau_2}) can be further simplified by assuming that
the temperature fluctuations behave as a  passive scalar (see Appendix~\ref{Spatial spectrum} and the discussion in Appendix~\ref{approx}). In that case,
it can been shown that the first term in the RHS of \eq{F_nu_tau_2}
vanishes \citep[see][Chap.~14.5]{MoninII}. Accordingly, \eq{F_nu_tau_2} simplifies as
\eqn{
{\cal F}_\tau (\tau,\nu) = {{ 72 \, (2\pi)^2  \,   {\kappa \rho} } \over R_s^2 } \,   \tilde{\cal B}_\Theta (\tau,\nu, k=0) \, , \label{F_nu_tau_3}
} 
with 
\begin{align}
\tilde{\cal B}_\Theta (\tau,\nu,0) & =  {1 \over \pi} \, \int \diff k \,
\left ( {{E_\Theta(k)}\over{k}} \right)^2 \, \psi_k(\tau,\nu) \, , \label{B_theta_2}\\
\psi_k(\tau,\nu) & = \int \diff \nu^\prime \, \chi_k(\nu^\prime) \, \chi_k(\nu^\prime+\nu) \, . \label{psi_k}
\end{align}
The final expression, \eq{F_nu_tau_3}, is recast to a more suitable form. To this end, we define  the characteristic wave-number $k_0=2\pi/\Lambda$ where $\Lambda$ is a characteristic length. We also define the characteristic frequency $\nu_0 = 1 / (2 \pi \, \tau_c)$ where $\tau_c$ is a characteristic time. Eventually, Eq.~(\ref{F_nu_tau_3}) leads  to  Eqs.~(\ref{F_nu_tau_4})-(\ref{N_g}) where  we have defined the dimensionless source function 
\begin{align}
S_\Theta (\tau,\nu) & =  {1 \over \pi} \, \int { {\diff K} \over {K^2} } \,{\tilde E}^2_\Theta(K) \, {\Psi}_K(\tau,\xi)\, , \label{source}
\end{align}
as well as the following dimensionless quantities:
\begin{align}
{\Psi}_K(\tau,\xi) & = \int \diff \xi^\prime \, {\tilde\chi}_K(\xi^\prime) \,  {\tilde\chi}_K(\xi^\prime +\xi) \, , \\
{\tilde E}_\Theta (K)& =  k_0 \, \Theta_{\rm rms}^{-2} \, E_\Theta(k) \, , \\
{\tilde\chi}_K (\xi) & = \nu_0 \, \chi_k(\nu)\, , \\
K & = \frac{k}{k_0} \, , \quad 
\xi = \frac{\nu}{\nu_0} \; ,
\label{K_xi}
\end{align}
where we have defined the characteristic wavenumber $k_0 = 2 \pi \, / \Lambda$, and the characteristic frequency $\nu_0 = 1 / (2 \pi \, \tau_c)$, where $\tau_c \equiv 1/ (k_0\, u_0)$ is a characteristic time and $u_0$ a characteristic velocity (see \eq{u0} below).
Note that the dimensionless quantities ${\tilde E} (K)$ and ${\tilde\chi}_K (\xi)$ verify the following normalization conditions:
\begin{align}
\int_{0}^{+\infty} \diff K \, {\tilde E}_\Theta (K) = { 1 \over 2} \, , \quad {\rm and} \quad
\int_{-\infty}^{+\infty} \diff \xi \, {\tilde\chi}_K (\xi) = 1 \, .
\end{align}

\subsection{Turbulence modeling}

\label{turbmdl}

\subsubsection{Spatial spectrum}
\label{Spatial spectrum}

As already mentioned, it is assumed that the temperature fluctuations behave as a passive scalar (see the discussion in Appendix~\ref{approx}). Therefore $E_\Theta(k)$ is given according to \citep[see][Chap VI-10]{Lesieur97}
\begin{align} 
E_\Theta (k) = \left \{ 
\begin{array}{lll} 
\displaystyle { \frac{a_0 \Theta_{\rm rms}^2} {u_0^2} \,  E(k)  } \, , & \textrm{for} & k \le k_c \\
\displaystyle {  \frac{a_0 \Theta_{\rm rms}^2 } {u_0^2} } \,  \left ( \frac{k_c} {k} \right )^4  \, E(k)  \, , & \textrm{for} & k > k_c \; ,
\end{array}
\right .
\label{eqn:E_s}
\end{align}
where $a_0$ is a normalization factor, $E(k)$ the kinetic energy spectrum, $u_0$ a characteristic velocity (see \eq{u0} below), and  $k_c$ the wavenumber, which separates two characteristic ranges:
\begin{itemize}
\item the inertial-convective range ($k<k_c$). In this domain, advection of the temperature fluctuations by the turbulent velocity field dominates over the diffusion. 
\item the inertial-conductive range ($k>k_c$). In this domain, the diffusion of the temperature  fluctuations dominates over advection.
\end{itemize}

We consider for $E(k)$  the Kolmogorov spectrum,  and the characteristic velocity $u_0$ is defined such that
\begin{align}
E(k)=  \left \{  
\begin{array}{lcl}
0 & \textrm{for} & k <0 \\
\displaystyle { \frac{u_0^2} {k_0} }  \,  K^{-5/3}  & \textrm{for} & k \ge k_0 \; ,
\label{eqn:kolmo}
\end{array}
\right .
\end{align}
where the characteristic velocity $u_0$ is defined such that
\eqn{
{3 \over 2} \, u_0^2 \equiv \int_0^{+\infty} \diff k \, E(k) \, .
\label{u0}
}

From \eq{eqn:E_s} and \eq{eqn:kolmo}, the spacial spectrum finally becomes 
\begin{align} 
{\tilde E}_\Theta (K) = \left \{ 
\begin{array}{lll} 
0 & \textrm{for} & K < 1 \\
\displaystyle {   a_0  \,  K^{-5/3}  }& \textrm{for} & 1 \le K \le K_C \\
\displaystyle {   a_0  \, K_C^4 \,  K^{-17/3} } & \textrm{for} & K_C<K  \; ,
\end{array}
\right .
\label{eqn:E_s_kolmo}
\end{align}
where we have defined $K_C= k_c/k_0$.

Some prescriptions are required for $k_c$  and $k_0$.
3D RHD show that from one stellar 3D model to another, the granule size $\Lambda$   scales approximately as the pressure scale-height ($H_p$) at the photosphere \citep{Freytag97}. Accordingly,   we assume that 
\begin{align}
k_0  =  \frac{2 \pi }{\Lambda} \,  \quad {\rm with} \quad
\Lambda  =  \beta \, H_p \; ,
\label{lambda_v}
\end{align}
where $\Lambda$ is a characteristic length-scale, $H_p$ the pressure scale height and $\beta$ is a free parameter.
Concerning the characteristic wavenumber $k_c$, in a medium with very low Prandtl number  (which is the case in the stellar medium),  $k_c$ is given by \citep[see][Chap.~VI]{Lesieur97}
\eqn{
k_c =  \zeta  \, \left ( { {\epsilon} \over  {\chi_{\rm rad}^3} }  \right )^{1/4} \; ,
\label{k_c}
}
where $\zeta$ is a free parameters introduced to exercise some control on this prescription, $\chi_{\rm rad}$ is the radiative diffusivity coefficient, and $\epsilon$ is the rate of injection of kinetic energy into the turbulent cascade. The latter is estimated according to $\epsilon = {1 \over 2} \, \Phi \, w^3 / \Lambda  $, where we have introduced the anisotropy factor $\Phi \equiv \left (    u_{\rm rms}^2 +  v_{\rm rms}^2  +  w_{\rm rms}^2  \right ) /  {w_{\rm rms}^2}$, where $u_{\rm rms}$, $v_{\rm rms}$, and $w_{\rm rms}$ are the rms of the  three components  of the velocity (horizontal and vertical ones).

\subsubsection{Frequency spectrum}
\label{Frequency spectrum}

Since in the inertial-convective range (\emph{i.e}, $k<k_c$) temperature fluctuations are dominated by advection, we  assume that -- at this scale range --  $\chi_k(\nu)$ is the same as the frequency spectrum associated with the velocity field, $\chi_k^{v}$. This hypothesis tends to be supported by large eddies simulations \citep[see e.g.][]{Samadi02II}.  Indeed, using a solar hydrodynamic simulation, \citet{Samadi02II} have found that this property is rather well verified by the entropy fluctuations. Since entropy fluctuations are mostly dominated by temperature fluctuations, this must be the same for the temperature fluctuations.
For the inertial-conductive range ($k>k_c$): temperature fluctuations are no longer dominated  by advection. However, to our knowledge,  no study has been conducted yet about the properties of  $\chi_k$ in this range. Therefore, we assume by default that in this range $\chi_k$ varies with $\nu$, as do $\chi_k^{v}$.

In a strongly turbulent medium,  $\chi_k^{v}$ is  well described by a Lorentzian function \citep{Sawford91,Samadi02II,Kevin10}, \emph{i.e.},
 \begin{equation} 
\chi_k(\nu) = {1 \over {\pi \nu_k}} \, \frac{\displaystyle{1}}{\displaystyle{1+\left( \nu/\nu_k \right)^2}} \; ,
\label{LF2}
  \end{equation} 
where $\nu_k$ is by definition the half-width at half-maximum of $\chi_k(\nu)$. In the framework of the \citet{Samadi00I} formalism, this latter quantity is evaluated as
\begin{align}
\label{nu_k}
\nu_k = \frac{k \, u_k}{2 \pi \lambda} \; ,
\end{align} 
with
\begin{align}
 u_k^2 = \int_{k}^{2k} E(k) \, {\rm d}k  \; ,
\end{align} 
where $\lambda$ is a free parameter introduced, following \citet{Balmforth92c}, to have some control on the adopted definition for $\nu_k$.
 We define the  characteristic time $\tau_c \equiv1/ (k_0 \, u_0)$, which corresponds to an estimate of the lifetime of the largest eddies. Accordingly, we have $\nu_0 = (2\pi \tau_c)^{-1}$. 

 The Lorentzian $\chi_k$ (\eq{LF2}) has a justification for a strongly turbulent medium. 
 However,   \citet{Georgobiani06}  have  found on the basis of a 3D RHD solar model that $\chi_k$  decreases more rapidly with $\nu$ near the photosphere than it does in deeper layers. 
Accordingly, as an alternative for a Lorenztian function and following \citet{Musielak94}, we also consider for  $\chi_k $   an exponential form  
 \begin{equation} 
\chi_k(\nu) = { {\ln 2 } \over {2 \, {\nu_k}}} \, \exp \left [ - \left | \ln 2 \, { \nu \over \nu_k } \right |  \right  ]  \, ,
\label{EF2}
  \end{equation}
where $\nu_k$ is the half-width at half-maximum.  We  alternatively adopted here the two different prescriptions for $\chi_k$. 


\section{Approximations and assumptions}
\label{approx}

Our theoretical model is based on three major  approximations and assumptions. They are discussed below.

{\noindent \it Passive scalar assumption:} It was assumed that  the temperature fluctuations behave as a passive scalar. We recall that a passive scalar $f$ is a quantity that
obeys an equation of diffusion \citep[e.g.,][]{Lesieur97}.
For the temperature fluctuations, this  equation of diffusion  is rigorously  valid when  the diffusion and the Boussinesq approximations are verified and the stratification is  negligible compared with the eddy size. 
However, all these conditions are not fulfilled in the vicinity of  the photosphere  where the granules are the more visible.  Indeed,  in this region the medium is optically thin such that the diffusion approximation does no longer hold.  Furthermore,  this region is characterized by a non-negligible turbulent Mach number that prevents the Boussinesq approximation from being  valid. Finally, the granule sizes are typically of the order of the pressure-scale height (see below). Nevertheless, as shown by \citet{Espagnet93} and \citet{Hirzberger97}s the spectrum, $E_\Theta$, associated with the temperature fluctuations at the surface of the Sun scales  with  $k$ as predicted by \eq{eqn:E_s}, which was derived assuming that the  temperature fluctuations behave as a passive scalar. This suggests that somehow the  temperature fluctuations obey  an equation of diffusion.

{\noindent \it Quasi-normal approximation:} The  approximation of Eq.~(\ref{qna}) assumes that fluctuating quantities are distributed according to a normal distribution. However, it is well known that the departure from the QNA is important in a 
strongly turbulent medium \citep[][]{Ogura63}. Furthermore, the upper-most part of the convection zone is a turbulent 
convective medium composed of essentially two flows (the granules and the downdraft plumes) that are asymmetric with respect to each other. Therefore, we obviously do not   deal with symmetric distribution, as it is the case for a normal distribution. With the help of a solar 3D hydrodynamical simulation, \citet{Kevin06a} have quantified the departure from the QNA \citep[seee also][]{Kupka07}.
 However, by comparing 3D hydrodynamical models representative of different main-sequence stars,  we have found that this departure does not vary significantly across the main sequence.

{\noindent \it Length-scale separation:} The derivation of \eq{F_nu_2} is based on the assumption that the product $\kappa \rho$ varies at a length-scale significantly longer than the granule size. However,  in the case of the Sun, for instance, the granules have a size of about 2~Mm \citep[see e.g.][]{Muller89,Roudier91} while the pressure-scale height, $H_p$, is of the order of few hundred kilometers at the photosphere.  Because near the photosphere the density scale-height $H_\rho$ is of the same order as $H_p$ and even lower, our  assumption does not hold near the photosphere. 
Nevertheless, 3D hydrodynamic models show that from a stellar model to another, the granule size scales as the pressure scale-height  at the photosphere \citep{Freytag97,Samadi08}. 
Therefore, as for the QNA, we expect that the departure from our hypothesis introduces a bias that remains almost constant across the HR diagram.  

The major approximations and assumptions adopted in our model are expected to be in default near the photosphere. However,  avoiding these approximations and assumptions would require additional theoretical improvements, and  they constitute --~ at the present time ~-- the only way for deriving an analytical model of the granulation spectrum.  
We also recall that our objective is to derive a simple analytical model for the interpretation of the observed scaling relations.
Furthermore, provided that the three free parameters involved in the  model are appropriately tuned, the theoretical model agrees reasonably well with the \citet{Ludwig06} 3D hydrodynamical approach (see Sect.~\ref{comp_direct}). 

\section{Eddy-time correlation }
\label{choice for chi}

Two different functions for $\chi_k(\nu)$ were tested: a Lorentzian function and an exponential one.
For a strongly turbulent medium, one expects a Lorentzian function \citep{Sawford91,Samadi02II,Kevin10}.
However,   with this function,  the theoretical granulation spectrum decreases as $\nu^{-2}$, while the observations decreases much more rapidly with $\nu$.
 On the other hand, adopting an exponential  $\chi_k$ results in a much better agreement  at high frequency.
This is because an exponential  $\chi_k$ decreases more rapidly with frequency than does a Lorenztian $\chi_k$. 
Finally, as in the Sun, adopting an  exponential  $\chi_k$ results in a much better match with the theoretical PDS computed with the ab~initio approach for two 3D hdyrodynamical models, one representative for an F-type star and the other one for a red giant star (see Sect.~\ref{comp_direct}).

 According to \citet[][see also \citet{Appourchaux10}]{Sawford91}, the time-Fourier transform of the \emph{Lagrangian} eddy-time correlation function is expected to tend to a Lorentzian function when the Reynolds number tends to infinity. In contrast, the less turbulent the medium (i.e., in general the lower Reynolds number), the more rapid the decrease of  $\chi_k$ with increasing $\nu$.  This behavior is also supported for the \emph{Eulerian} eddy-time correlation ($\chi_k$)  by hydrodynamical numerical models \citep[see e.g.][]{Samadi10}.
In other words, the $\nu$ variation of  $\chi_k$ is expected to depend on the degree of turbulence.
Accordingly, our result  confirms that the granules, which are mainly visible near the photosphere,  are  less turbulent than the super-adiabatic layers situated a few hundred kilometers below the photosphere. 
We therefore confirm the previous claim by  \citet{Nordlund97} that the granules have a low level of turbulence. This is also consistent with the results by \citet[][see also \citet{Nordlund09}]{Georgobiani06}. Indeed, these authors showed that $\chi_k$  decreases more rapidly with $\nu$ near the surface of a solar 3D model than it does in deeper  layers, where $\chi_k$ is close to a Lorentzian function \citep{Samadi02II}. 

Finally, the fact that the granules have a relatively  low level of turbulence is probably not specific to the Sun. Indeed, the 3D RHD models of stars show that the granules have similar properties as in the Sun \citep[e.g.,][]{Trampedach13}. Therefore, it is not surprising that the theoretical PDS computed with the ab~initio approach varies with $\nu$ in a similar way as in the Sun.



\section{Remaining discrepancies}
\label{residual}

 Although globally satisfactory,  the theoretical model does not  reproduce  the solar granulation spectrum perfectly. Indeed, the observed spectrum shows a kink at $\nu \sim 1$~mHz (see Sect.~\ref{LF_vs_EF}). 
There is no consensus yet about the physical origins of this feature.   Depending on the authors, it  is either attributed to the occurrence of bright points \citep[e.g.,][]{Harvey93,Aigrain04},  the changing properties of the granules \citep{Andersen98}, a  second granulation population \citep{Vazquez-Ramio05}, and finally to faculae \citep{Karoff11}.
A similar discrepancy was obtained by \citet{Ludwig09} on the basis of the {\it ab~initio} approach. This indicates that pure hydrodynamical approaches cannot fully account for the observed solar granulation spectrum. 
Nevertheless, the remaining discrepancies  represent only a small fraction of the total brightness fluctuations produced by the granulation phenomenon. 

The theoretical  model does not perfectly reproduced   the PDS obtained with the {\it ab~initio} approach for the two 3D models considered in this work. For instance, a difference of less than about  15\,\%  is obtained for $\sigma$ with the 3D model for an F-dwarf star and a  difference less than about 30\,\% are obtained for $\taueff$ with the 3D model for a red giant star. 
These differences must   be attributed to the different approximations and assumptions adopted (see Appendix~\ref{approx} above). Nevertheless, they remain  of the order of the dispersion obtained between the different methods of analysis investigated by  \citet[][see also paper~II]{Mathur11}.
Because  the {\it ab~initio} modeling  is based on a very limited set of physical hypothesis, the reasonable  agreement between the two approaches shows that despite  the numerous  hypotheses and assumptions adopted in our theoretical model (see Appendix~\ref{approx}), our model provides realistic results.

\end{document}